\documentclass{emulateapj}
\usepackage{url}
\usepackage{color}
\usepackage{latexsym, graphicx, amssymb, longtable, epsf}
\newcommand{\x}{}
\newcommand{\rev}[1]{{#1}}

\long\def\symbolfootnote[#1]#2{\begingroup\def\thefootnote{\fnsymbol{footnote}}
\footnote[#1]{#2}\endgroup}

\shorttitle{24 new planets in 12 systems}
\shortauthors{Xie}

\begin{document}

\title{Transit Timing Variation of Near-Resonance Planetary Pairs:  Confirmation of Twelve Multiple Planet Systems}

\author{Ji-Wei Xie$^{1, 2}$}
\affil{$^1$Department of Astronomy \& Key Laboratory of Modern Astronomy and Astrophysics in Ministry of Education, Nanjing University, 210093, China; Xiejiwei@gmail.com}
\affil{$^2$Department of Astronomy and Astrophysics, University of Toronto, Toronto, ON M5S 3H4, Canada; jwxie@astro.utoronto.ca}


\begin{abstract}
We extract Transit Timing Variation (TTV) signals for 12 pairs of transiting planet candidates that are near first-order Mean Motion Resonances (MMR), using publicly available
 \emph{Kepler} light curves (Q0-Q14).  These pairs show significant sinusoidal TTVs with theoretically predicted periods, which demonstrate these planet candidates are orbiting and interacting in the same system.  Although individual masses cannot be accurately extracted based only on TTVs because of the well known degeneracy between mass and eccentricity, TTV phases and amplitudes can still place upper limits on the masses of  the candidates, confirming their planetary nature. Furthermore, the mass ratios of these planet pairs can be relatively tight constrained using these TTVs. The planetary pair in Kepler-82 (KOI-880) seems to have a particularly high mass ratio and density ratio, which might indicate very different internal compositions of these two planets. Some of these newly confirmed planets are also near MMR with other candidates in the system, forming unique resonance chains, e.g.,  Kepler-80 (KOI-500).

\end{abstract}



\section{Introduction}
As of today, the \emph{Kepler mission} has found more than 2700 planet candidates \citep{Bat12, OD12, Hua12, Bur13}, and more than 880 of them are multiple transiting candidates  \citep{Fab12b}. Of particular interest in these multiple transiting systems is the study of Transit Timing Variations (TTVs), where the time of transit deviates from strict Keplerian periodicity. \x{One natural cause of TTVs  could be the gravitational interaction with other planets}. By comparing dynamical models to the TTV data, one can confirm the existence of planets and/or even constrain their masses and orbits properties \citep{HM05, Ago05}. Recently, thanks to the TTV technique, many Kepler planet systems have been confirmed and characterized \citep{Hol10, Lis11, Coc11, For12a, Ste12, Fab12a, Nes12}. Most of these planetary systems are near Mean Motion Resonance (MMR). This is expected as planets in MMR build up their mutual interactions and thus induce particularly large TTV signals.  

One of the keys to characterizing planets using TTVs is to find a good TTV model that can fit the TTV data.  To do this, most of the above studies rely on N-body simulations to calculate the model TTVs \citep{Ver11}, which is computationally expensive, and thus it is not practical for characterizing the hundreds of Kepler candidates. 
On the other hand, recently, \citet{Lit12} (Paper I hereafter) derived new formulae for the TTVs from two near-resonance planets, and proposed a new method to analyze TTV data. Compared to the method of N-body simulations, the new method is much less time-consuming (hence suitable for analysis of a large sample), and it is straightforward to obtain the mass and eccentricity information of the system.

In this paper, first, in section 2, we measure the TTVs of 12 new KOI\footnote{During the revision process of this paper, several systems reported here have been independently confirmed by \citet{Rag12, Ste13} } pairs near first order of MMR. Then, in section 3, applying the new method (Paper I), we analyze the TTV of each system, which allows us to confirm their planetary nature. We finally conclude in section 4 with comments on some particularly interesting systems.

\section{TTV Measurement }

The data used in this paper are the long cadence (LC), ``corrected'' light curves (PDC) of KOIs from Q0 to Q14, which are available at Multi-mission Archive at STScI (MAST{\footnote{http://archive.stsci.edu/kepler}}). 

\begin{figure}
\begin{center}
\includegraphics[width=\textwidth]{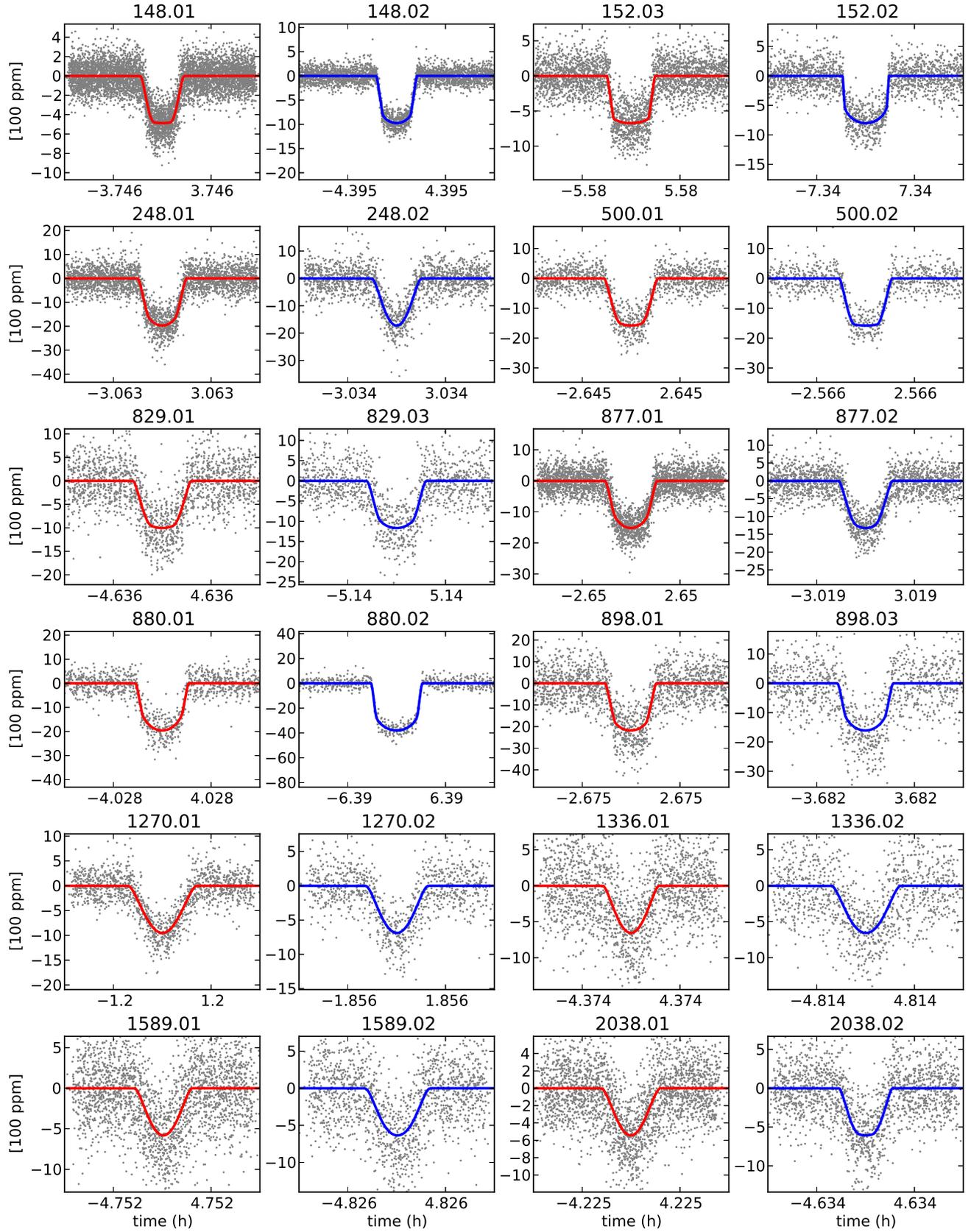}
\caption{\rev{Folded light curves (superposition of all transit segments, setting all central transit times = 0) of 12 pairs of KOIs. On the top of the light curves (solid points), there are red and green solid lines showing the best transit model fits to the inner and outer one of each pair, respectively. Note the different scales in the horizontal and vertical axes}}
\label{fig_temp}
   \end{center}
\end{figure}

For each planet candidate, we measure two sets of transit times. 
We obtain the first set by performing several iterations as described below.
\begin{itemize}
\item \emph{Step 1. Segments.} We estimate the mid-transiting times ($t_{\rm mid}$) using the linear ephemerides based on the epoch and period reported in the NASA exoplanet archive \footnote{http://exoplanetarchive.ipac.caltech.edu}. Surrounding each mid-transit time, we cut off a segment of light curve with length of 4 times the transit duration ($t_{\rm dur}$). Segments which have missing data due to gaps in light curves or overlap with segments of other planet candidates by more than one transit duration are removed. All the selected segments are de-trended using a \rev{polynomial} base line\rev{\footnote{\rev{We tried different orders (up to 4) for the polynomial base line, and select the one with the lowest AICs score \citep{BA02}, \rev{namely $AICs =\chi^{2}+2nk/(n-k-1)$, where $\chi^{2}$ and $k$ are the chi square of the best fit and number of free parameters of the base line respectively, and $n$ is the number of data points in a segment.  In most cases, we find that a linear base line is the best choice.}}}}, and their fluxes are normalized to unity. 
 \item \emph{Step 2. Template.} Segments from the last step are then superimposed to form the folded light curves and are fitted as a whole to form a template using the theoretical transiting model \citep{MA02}.  During the fitting, the epoch, planet-star radius ratio, transit duration, impact parameter, and linear limb-darkening coefficient are allowed to vary. Figure \ref{fig_temp} plots the folded light curves and the template fits for the 12 KOI pairs studied in this paper. 
 \item \emph{Step 3. Transit Times.} Segments are fitted individually to the template. Following \citet{For12a}, only the mid-transit time is allowed to vary and all other parameters are fixed on the values given by the last template. The transit time is determined by Levenberg-Marquardt (LM)\rev{\footnote{\rev{To reduce the probability that Levenberg-Marquardt method becomes trapped in a local minima (if any), we tried 5 different initial mid-transit time ($t_{\rm mid}-t_{\rm dur}$, $t_{\rm mid}-0.5t_{\rm dur}$, $t_{\rm mid}$, $t_{\rm mid}+0.5t_{\rm dur}$ and $t_{\rm mid}+t_{\rm dur}$), and select the best fit with the lowest $\chi^{2}$. }}} minimization of $\chi^{2}$, and its uncertainty is estimated from the covariance matrix \citep{Pre92, Mar09}. The updated transit times are then used as the input in step 1 for the next iteration, and they converge generally in two to three iterations.
 \end{itemize}
 
 \rev{The above transit time measuring method is fast as there is only one parameter to float when using a fixed template in step 3. Similar methods have been widely used to confirm and characterize many Kepler planets with TTV \citep{Hol10, Lis11, Coc11, For12a, Ste12, Fab12a, Lis13}. However, using a fixed template brings some concern that it ignores the uncertainties and  possible variations of other basic transit parameters (e.g., transit duration variation \citep{Nes13}), which, in principle, could affect the transit  time measurement. }

\rev{For the above reason, we further consider a second set of transit time measurements as the follows. Following \citet{Nes13}, we fit every transit (the segments after the above the Step 3) individually by letting all transit parameters (mid-transiting time, transit duration, impact parameter, planet-star radius ratio and linear limb-darkening coefficient) vary and reduce the binning effect of LC data using the resample method as in \citet{Kip10}. This process uses a Markov Chain Monte Carlo (MCMC) method \citep{Pat10}, in which we adopt the fitting results from the above steps 2 and 3 as the first guesses of the proposal distributions of those fitted parameters. These proposal distributions are then automatically adjusted a few times until the acceptance rates reach a range of 20-50$\%$. For each transit fitting, we run 5 MCMC chains and perform the \citet{GR92} test to ensure that they converge. \rev{\footnote{Specifically, we discard results with Gelman-Rubin statistic larger than 1.1}}. Each chain has a length of 120,000 states and the \rev{first 20, 000 ones are abandoned for burn-in. To reduce the correlation of consecutive states, we adopt a thinning interval of 50 (ten times the number of free parameters \citep{For05}), finally leaving  us 10, 000 (2000x5) states in total for each fit, whose median and standard deviation are used as the best fit and corresponding one-sigma uncertainty. The drawback of the full MCMC fitting used here is that it is over 100 times slower than the LM fitting used in measuring the first set of transit times. Typically, the MCMC fitting takes 3-6 hours for each transit segment. As there are thousands of segments to fit for the 12 multiple systems studied in this paper, we had to let over 100 CPUs work in parallel to finish all the fitting jobs in several days. }}

\begin{figure}
\begin{center}
\includegraphics[width=\textwidth]{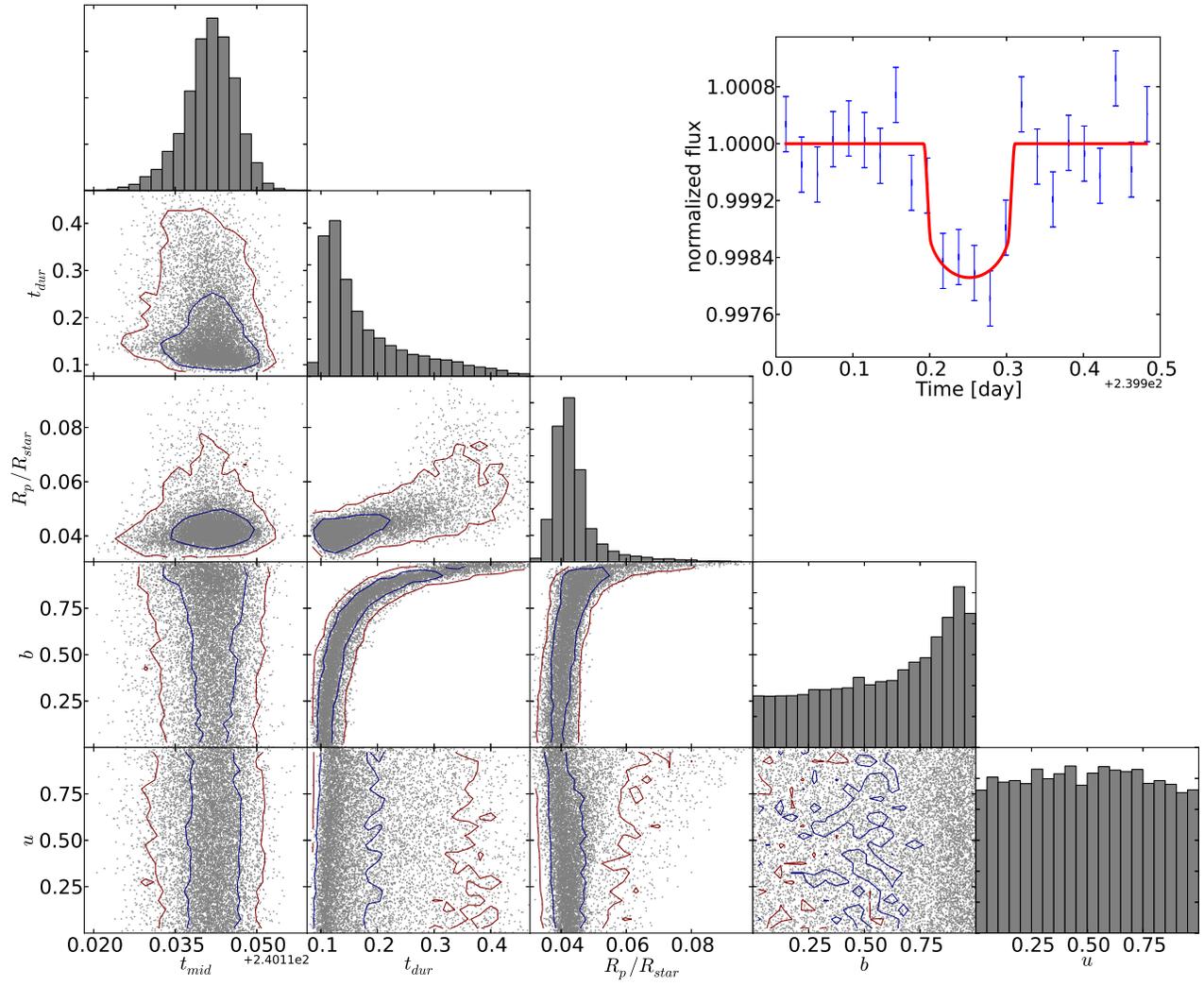}
\caption{\rev{MCMC fitting results of a randomly chosen transit for KOI 248.01.  The triangle plot in the bottom left shows the 1-D and 2-D distributions of 10,000 states for each fitted parameter. The blue and red lines indicate the regions with confidence levels of 0.683 and 0.955, respectively. The top right panel shows the best transit fit model (using median of the 10,000 states) on the top of the light curve data points. We see that the mid-transit time is nearly uncorrelated with any other transit parameters, but the transit duration, impact parameter and planet-star radial ratio are correlated with each other. Furthermore, the impact parameter and linear limb-darkening coefÞcient are poorly constrained. These results are typical of all individual transits studied in this paper. } }
\label{fig_mcplot}
   \end{center}
\end{figure}

\begin{figure}
\begin{center}
\includegraphics[width=\textwidth]{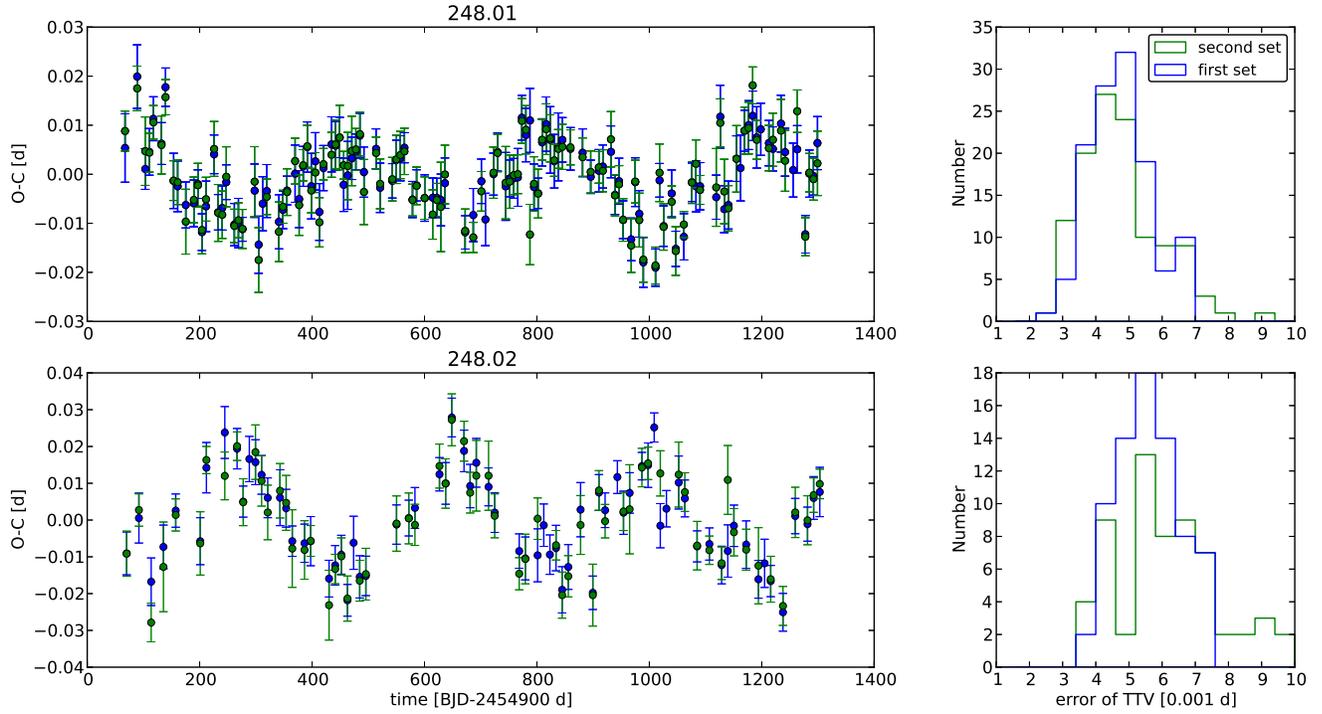}
\caption{\rev{Comparison of two set of TTVs. The left two panels plot the first set of TTV (blue, using LM fitting with a fixed template) and the second set of TTV (green, using MCMC fitting) for KOI 248.01 and 248.02, respectively. The right two panels plot the error bar histogram of these two set of TTVs.  As can be seen, these two set of TTVs are consistent with each other, although there are notably larger error bars for the second set.}}
\label{fig_comp}
   \end{center}
\end{figure}

\rev{Figure \ref{fig_mcplot} plots the MCMC fitting results for a randomly chosen transit for KOI-248.01, which shows some typical behaviours we see from all individual transits studied in this paper. First, the impact parameter and limb darkening coefficient are poorly constrained. Second, transit duration, impact parameter and planet-star radial ratio are correlated with each other. Last but not least, the mid-transiting time is nearly uncorrelated with  any other transit parameter. Such an independence of mid-transit time is consistent with the analytical results of \citet{Car08}\footnote{\rev{In the case when the limb darkening effect can be ignored,\citet{Car08} analytically derives that the mid-transit time is an independent parameter (not correlated with any other basic transit parameter). Figure \ref{fig_mcplot} shows that such an independence is nearly unaffected even if the limb darkening effect is considered.}}, which somewhat supports the previous approximation of using a fixed template to measure the mid-transiting time.  Indeed, comparison between the two sets of transit times (see Fig.\ref{fig_comp} for an example) shows they are completely consistent with each other,  although there are notably larger error bars for the second set. However, considering the expensive computing cost for the MCMC fitting, we suggest using the LM method to provide a bird's eye view of the TTV of a large KOI sample and using MCMC fitting for detailed characterization of individual KOIs, e.g., this work. }
  
\rev{Note, for both sets of transit time measurements we did not fit the time correlated noise. To address this issue, we perform a further test for the residuals from the above MCMC fitting. Following \citet{Pon06} we calculate the variances of gradually increased binned residuals and fit them as $V(n) = \sigma_{w}^{2}/n+\sigma_{r}^{2}$, where $n$ is the number of data points in a bin, $\sigma_{w}$ indicates the white noise and $\sigma_{r}$ the red noise. To ensure that our results are not significantly affected by red noise, we exclude a handful of transit time measurements with $\sigma_{r}/\sigma_{w}>0.1$.} 

\rev{For the remaining transit time measurements, we throw away outliers, i.e., those with an absolute deviation from the linear ephemeris exceeding four times the median absolute deviation of transit times from the linear ephemeris, or those with a transit time uncertainties exceeding twice the median of the transit time uncertainties of the candidate. We use the second set of TTVs (table 1) for the further analyses, although these two sets of TTVs lead to similar results.}

 
 \section{TTV Analysis}
 In this paper, we focus on planet candidate pairs which are near first order MMR, 
i.e, 
\begin{eqnarray}
\frac{P^{'}}  {P}  \sim \frac{j}{j-1}
\end{eqnarray}
where $j=2,3, . . .$,  $P$ and $P^{'}$ are the orbital periods of the inner and outer candidates, respectively.
Through out this paper, we adopt such a convention: properties of the outer (inner) candidate in a pair are denoted with (without) a superscript  ``$'$ ''.
The proximity to resonance is defined as in Paper I:
\begin{eqnarray}
\Delta = \frac{P^{'}}  {P} \frac{j-1}{j} -1 .
\end{eqnarray}

Our method of analyzing TTV near resonance has been justified and presented in Paper I. \footnote{\x{This method is valid for planets pair \emph{near} MMR, i.e., $\Delta>\sqrt{\mu e_{\rm free}}$, where $\mu$ is the planet-star mass ratio and $e_{\rm free}$ is the free eccentricity (see the end of Appendix in Paper I). This requirement is usually satisfied given $|\Delta|\sim0.01$ and $e_{\rm free}\sim0.01$ on order of magnitude for typical Kepler planets \citep{WL12}}}.    Here, we briefly summarize it and describe the specific procedure used in this paper.  
  
 \subsection{TTV: Model Fitting}
 As derived in Paper I, the transit time series of two transiting objects near first order MMR are two linear+sinusoidal curves in the following forms. 
 \begin{eqnarray}
 t = T + P \times n + |V|{\rm sin}(\lambda^{j} + \angle V), \nonumber \\
 t^{'} = T^{'} + P^{'} \times n^{'} + |V^{'}|{\rm sin}(\lambda^{j} + \angle V^{'}),
 \label{ttveqn}
\end{eqnarray}
where $t$ and $t^{'}$,  $n$ and $n^{'}$, $T$ and  $T^{'}$, $P$ and $P^{'}$ and  $V$ and $V^{'}$ are the transit time,  transit sequence number, transit epoch, transit period, and TTV complex (see Eqn.7, $\angle V$ denotes the phase of complex $V$) of the inner and outer planets, respectively. And 
\begin{eqnarray}
\lambda^{j} & = & j\lambda^{'}-(j-1)\lambda \nonumber \\
                     & = & -\frac{2\pi}{P^{'}/j\Delta} (t-T^{'}) + \frac{2\pi}{P} (j-1)(T-T^{'})
\end{eqnarray}
is the \emph{longitude of conjunctions}, where $\lambda = \frac{2\pi}{P}(t-T)$ and $\lambda^{'} = \frac{2\pi}{P^{'}}(t-T^{'})$ are the mean longitudes of the inner and outer planet, respectively. The TTV period, which we refer to as the \emph{super-period}, is 
\begin{eqnarray}
P^{j}=\frac{P^{'}}{j|\Delta|},
\label{Pj}
\end{eqnarray}
and 
\begin{eqnarray}
A_{\rm ttv}  =   |V|, \,\,  & A^{'}_{\rm ttv}  =   |V^{'}| \nonumber \\
\phi_{\rm ttv} =  \angle \left(\frac{\Delta}{|\Delta|}V\right), \,\, &  \phi^{'}_{\rm ttv} = \angle \left(\frac{\Delta}{|\Delta|}V^{'}\right)
\label{amp_pha}
\end{eqnarray}
are the TTV amplitudes and phases, respectively.

\x{We use a MCMC method \citep{Pat10} to fit above model to the TTV data.}  
Figure \ref{fig_ttv1} and \ref{fig_ttv2} plot the TTV data and best fits for the 12 KOI pairs. The fitting results are summarized in table 2.

\subsection{TTV: FAP Analyses}
Figure \ref{fig_ttv1} and \ref{fig_ttv2} illustrate 12 KOI pairs, each showing sinusoidal TTVs with theoretically predicted period (super-period). To address how likely that such a pair of TTVs could be produced by coincidence due to  data noise rather than a real signal of a pair interacting planets, we perform the following two sets of false alarm probability (FAP) analyses. 

The first set of FAP, i.e., $\rm FAP_{1}$, is based on the Lomb-Scargle (LS) periodogram \citep{Sca82, ZK09}. Specifically,  for each pair of measured TTV, we first compute the LS periodogram and record their powers at the super-period. We then calculate the LS powers at the super-period for another $10^{5}$ sets of random permutations of the original TTV data.  $\rm FAP_{1}$ is estimated as the fraction of realization pairs with both LS powers larger than that of the corresponding original TTV.   

The second set of FAP, i.e., $\rm FAP_{2}$, is based on refitting $10^{4}$ sets of random permutations of the original TTV data using the procedure mentioned in section 3.1 \x{but with the LM fitting algorithm (because MCMC is too time consuming to run $10^{4}$ fittings to each KOI pair)}.    $\rm FAP_{2}$ is estimated as the fraction of realization pairs with both fitted TTV amplitudes larger than the corresponding original TTV amplitudes. This refitting method is similar to that used in \citet{Fab12a}.

As an example of the results, figure \ref{fig_fap} plots the FAP analysis for KOI pair 248.01 and 248.02. The full results ($\rm FAP_{1}$ and $\rm FAP_{2}$) are listed in table 2. All of them are well below $10^{-3}$, leading to very high confidence on the observed TTV pairs, which demonstrates that each pair of transiting objects are really orbiting and interacting in the same system. Note, the FAP here is a system statistic for a pair KOIs rather than for individual KOIs. A low FAP does not require both transiting objects to have very strong clear TTVs as KOI-248.01(02). A weak  noisy TTV, e.g., the one of KOI-1589.01,  can also be confidently detected if its partner clearly shows the predicted TTV, e.g, KOI-1589.02.

\subsection{TTV: Constraints on Planet Mass and Mass Ratio}
The advantage of this TTV analysis method is that the TTV amplitudes and phases (or TTV complex) explicitly reveal the masses and eccentricity of the system (Paper I), i.e., 
\begin{eqnarray}
V  & = &  P \frac{\mu^{'}}{\pi j^{2/3}(j-1)^{1/3}\Delta}\left(-f-\frac{3}{2}\frac{Z^{*}_{\rm free}}{\Delta}\right) \nonumber \\
V^{'}  & = &  P^{'} \frac{\mu}{\pi j\Delta}\left(-g+\frac{3}{2}\frac{Z^{*}_{\rm free}}{\Delta} \right),
\label{vv}
\end{eqnarray}
where $\mu$ and $\mu^{'}$ are the mass ratio of the inner and outer objects to the star, respectively, 
$f$ and $g$ are sums of Laplace coefficients (of order of unity), as listed in Table A1 of Paper I, and $Z^{*}_{\rm free}$ is the complex conjugate of 
$Z_{\rm free}=f z_{\rm free} + g z^{'}_{\rm free}$, a linear combination of the free complex eccentricities of the two planets.  For typical Kepler systems considered here, $\mu\le10^{-4}$, $\mu^{'}\le10^{-4}$, and $|\Delta|\sim10^{-2}$, the forced eccentricities of the two planets are relative small, i.e., $|z_{\rm forced}|\sim \mu/|\Delta|\le10^{-2}$ and $|z^{'}_{\rm forced}|\sim \mu^{'}/|\Delta|\le10^{-2}$ . Thus, taking an approximation of order of unity,  $|Z^{*}_{\rm free}|$ in Eq.\ref{vv} could be considered as the true eccentricity of the two planets. 

In principle,  $\mu$, $\mu^{'}$ and $Z^{*}_{\rm free}$ (both real and imaginary parts) can be inferred by inverting Eqn.\ref{vv}. In reality, however, $Z^{*}_{\rm free}$ is highly degenerate with $\mu$ and/or $\mu^{'}$ (see below and also in Paper I). Nevertheless, as mentioned in Paper I,  one can derive the nominal masses from Eqn.\ref{vv} by assuming $|Z^{*}_{\rm free}|=0$, i.e.,
\begin{eqnarray}
m_{\rm nom} &=& M_{\star}\left|\frac{V^{'}\Delta}{P^{'}g}\right|\pi j \nonumber \\
m^{'}_{\rm nom}& =& M_{\star}\left|\frac{V\Delta}{Pf}\right|\pi j^{2/3}(j-1)^{1/3},
\label{mn}
\end{eqnarray}
As shown in figure 10 of Paper I, statistically, planets' true masses are likely less than the nominal masses. 
However, one should note that the nominal mass is not the upper limit. In order to further derive the upper mass limit from the TTV amplitude and phase for each planet candidate, we perform some Monte Carlo simulations via the following steps.


\begin{itemize}
\item \emph{Step 1}. We generate a $Z_{\rm free}$ prior assuming its phase is from a uniform distribution from $0$ to $2\pi$, and its modulus from some certain distributions (see below).

\item \emph{Step 2}. Using above $Z_{\rm free}$ and Eqn.\ref{vv}, we calculate the corresponding TTV phases, $\phi$ and $\phi^{'}$.

\item \emph{Step 3}. We calculate how likely (probability) the above phases could be the observed TTV phases using,
\begin{eqnarray}
{\rm Prob}  & = &  \frac{1}{\sigma\sqrt{2\pi}} \rm exp\left(-\frac{(\phi-\phi_{ttv})^{2}}{2\sigma}\right),\nonumber \\
{\rm Prob^{'} } & = &  \frac{1}{\sigma^{}{'}\sqrt{2\pi}} \rm exp\left(-\frac{(\phi^{'}-\phi_{ttv})^{2}}
{2\sigma^{'}}\right),
\label{prob}
\end{eqnarray}
where we have assumed the measured TTV phase (table 2) is a gaussian distribution with a centroid of $\phi_{\rm ttv}$ ($\phi_{\rm ttv}^{'}$) and one-sigma uncertainty of $\sigma$ ($\sigma_{'}$).

 \item \emph{Step 4}. Two random numbers $c$ and ${c'}$ are drawn from a uniform distribution between 0 and 1. We then  take the $Z_{\rm free}$ generated at step 1 as one realization to its posterior distribution if ${\rm Prob} \ge c$ and ${\rm Prob^{'}} \ge c'$, otherwise we go back to step 1.
 
 \item \emph{Step 5}.  We generate a pair of TTV amplitudes $|V|$ and $|V^{'}|$ from Gaussian distributions with the centroids and deviations equal to the measured TTV amplitudes and their uncertainties (table 2). 
 
 \item \emph{Step 6}. Using the TTV amplitudes, $|V|$ and $|V^{'}|$,  and the free eccentricity, $Z_{\rm free}$,  obtained in step 4 and 5, the two planets' masses can be solved out from Eqn.\ref{vv} if assuming a stellar mass (table 3). Such two derived planets' masses ($m$ and $m^{'}$) are one realization to their posterior distributions. 
 \end{itemize}
 The above steps are repeated for 1000 realizations of $Z_{\rm free}$, $m$ and $m^{'}$.
 
 For each KOI pair, we perform ten sets of the above simulations with different prior distributions of the modulus of free eccentricity, $|Z_{\rm free}|$. Five of them assume a uniform distribution of $|Z_{\rm free}|$ between 0 and five different upper cutoffs (0.4, 0.2, 0.1, 0.05 and 0.025). The other five assume a Rayleigh distribution of $|Z_{\rm free}|$ with different means (0.2, 0.1, 0.05, 0.025 and 0.0125). We adopt the maximum mass of these 10 set of $m$ ($m^{'}$) as the planet's maximum mass (table 3), $m_{\rm max}$($m^{'}_{\rm max}$).  We focus on free eccentricity below 0.4 since previous studies on TTV (e.g., paper I) and on transit duration (Moore et al. 2011) suggest most KOIs, especially those in multiple transiting systems, are likely have low eccentricities. Extending $|Z_{\rm free}|$ prior to higher values generally leads to a smaller $m_{\rm max}$($m^{'}_{\rm max}$).
 
 Figure \ref{fig_mass1} plots the results of three sets of above simulations for KOI-248.01 and KOI-248.02. As expected, the masses are strongly correlated with eccentricity due to the well-known degeneracy between them, and thus generally, one cannot accurately extract individual masses or eccentricities based solely on their TTV \footnote{Sometimes the TTV phases and/or their distribution may help break the degeneracy (see paper I and \citet{WL12})}.    The posterior mass distribution has a narrow high-end tail, which shows that extremely high mass is possible only in a very narrow parameter space, i.e., the $Z_{\rm free}\sim (-2/3)f\Delta$ or $Z_{\rm free}\sim (2/3)g\Delta$ as seen from Eqn.\ref{vv}. However, none of our 1000 realizations here reaches a mass larger than 100 $M_{\rm E}$, which shows that TTVs can still place a strong constraint on the mass range, and KOI-248.01 and KOI-248.02 are thus confirmed to be planets rather than brown dwarfs. Furthermore,  as can be seen from the right column of figure \ref{fig_mass1}, TTVs also place a relatively tight constraint on the mass ratio of the KOI pair. 
 Similar results can also be seen in figure \ref{fig_mass2} and \ref{fig_mr}, which plot the results of the ten sets of simulations for the 12 KOI pairs. 
 As can be seen, all the 12 KOI pairs (24 candidates) have a maximum mass less than 25 Jupiter masses ($M_{\rm J}$) or $< 7945$ Earth masses ($M_{\rm E}$), confirming their planetary nature (Schneider et al. 2011).

\section{Discussions and Conclusions}
\subsection{24 confirmed planets in 12 planetary systems}
In this paper, we measure and analyze the TTVs of 12 KOI pairs. All of them show theoretically predicted  sinusoidal TTV pairs with very high confidence (Fig.\ref{fig_ttv1}, \ref{fig_ttv2}, and table 2), and their TTV phases and amplitudes constrain their masses within the planetary range, leading to the confirmation of 24 planets in 12 planetary systems (table 3). Although such a sinusoidal TTV is a strong evidence of a pair of interacting planets against other astrophysical false positives\footnote{Such as two planets orbiting two different background stars or planets orbiting in a binary system, as discussed in \citet{Ste13}}, we perform two more checks as the following to further ensure our confirmation of these planet systems.

First, we check the centroid offsets of these targets during transit. If a pair of planets were orbiting different stars, then each planet transit would cause some offsets to the target centroid, and the offsets caused by different planets would be in different directions. In all cases studied here, we do not find any significant offset (personal communication with Ji Wang), which is consistent with each pair of planets orbiting the same star. 

Second, we calculate the normalized transit duration ratio, $\xi=(T_{\rm dur}/T_{\rm dur}^{'})(P^{'}/P)^{1/3}$ \citep{Fab12b}, for each pair. The value of $\xi$ should be order of unity if the pair planets orbit the same star. Indeed, in all cases as listed in table 3,  it is consistent with each pair of planets being in the same system.

In figure \ref{fig_sys}, we plot these 12 systems according to orbital periods and radii. Most of these systems (except for KOI-1270) are multiple systems with 3-5 transiting planet candidates. The two planets confirmed using TTVs in each system are usually the largest ones (except for KOI-152, see below for more discussions), which is consistent with our expectations as TTV amplitudes and their measurement accuracy generally increase with planet size (mass). We discuss some interesting systems below.

\subsection{Comments on some individual systems}
\begin{figure}
\begin{center}
\includegraphics[width=\textwidth]{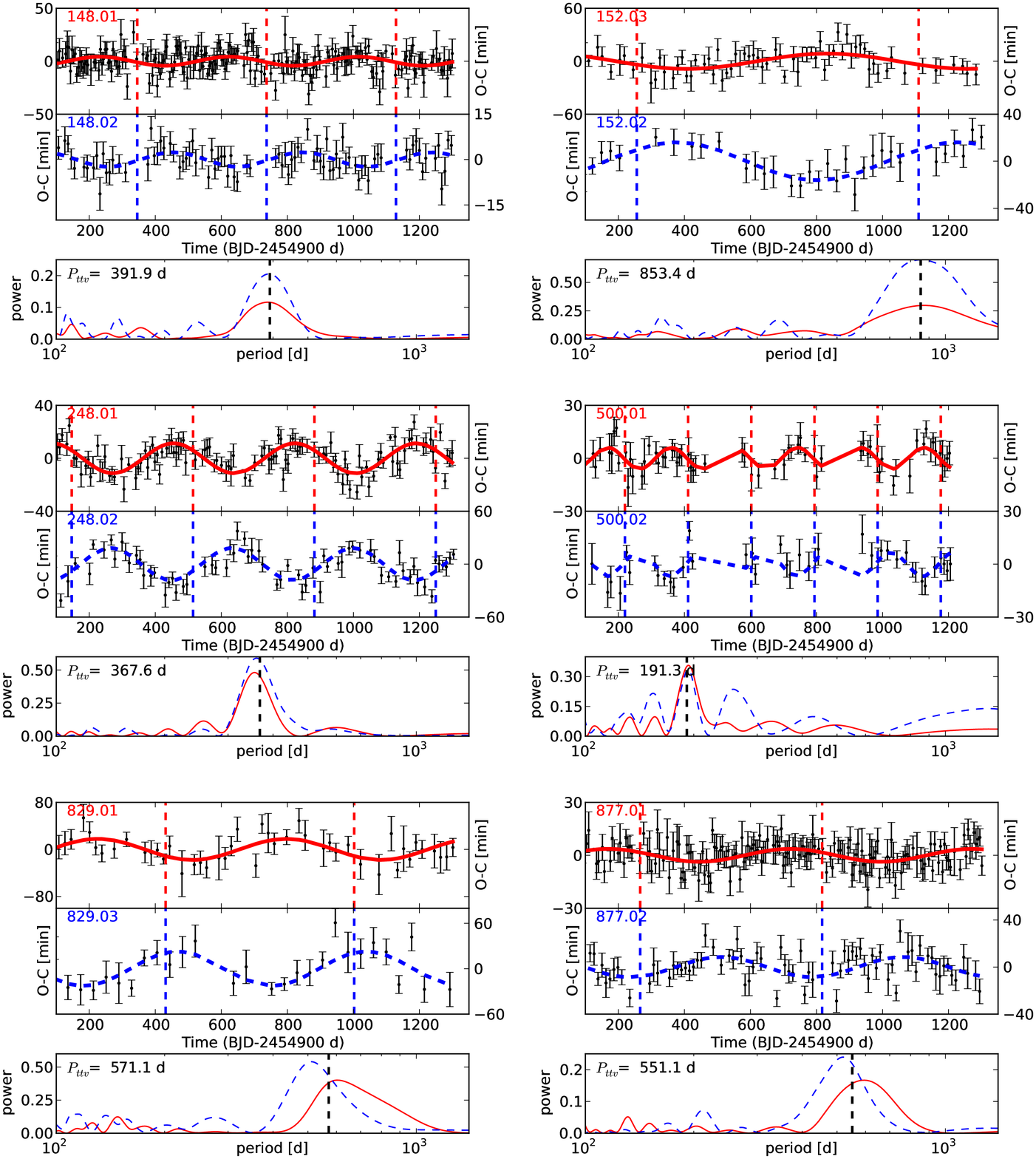}
\caption{\rev{TTV evidences for six KOI pairs (red lines for the inner one, and blue for the outer one). For each of them, we plot the best-fit theoretical  curves on top of the TTV data and the TTV periodogram.  In the TTV fitting panels, the vertical dashed lines  denote the times when the longitude of conjunction points at the observer, i.e., $\lambda^{j}=0$. In the periodogram panel, the vertical dashed line denotes the theoretically predicted period of the TTV. The TTV fitting results are summarized in Table 2.}}
\label{fig_ttv1}
   \end{center}
\end{figure}

\begin{figure}
\begin{center}
\includegraphics[width=\textwidth]{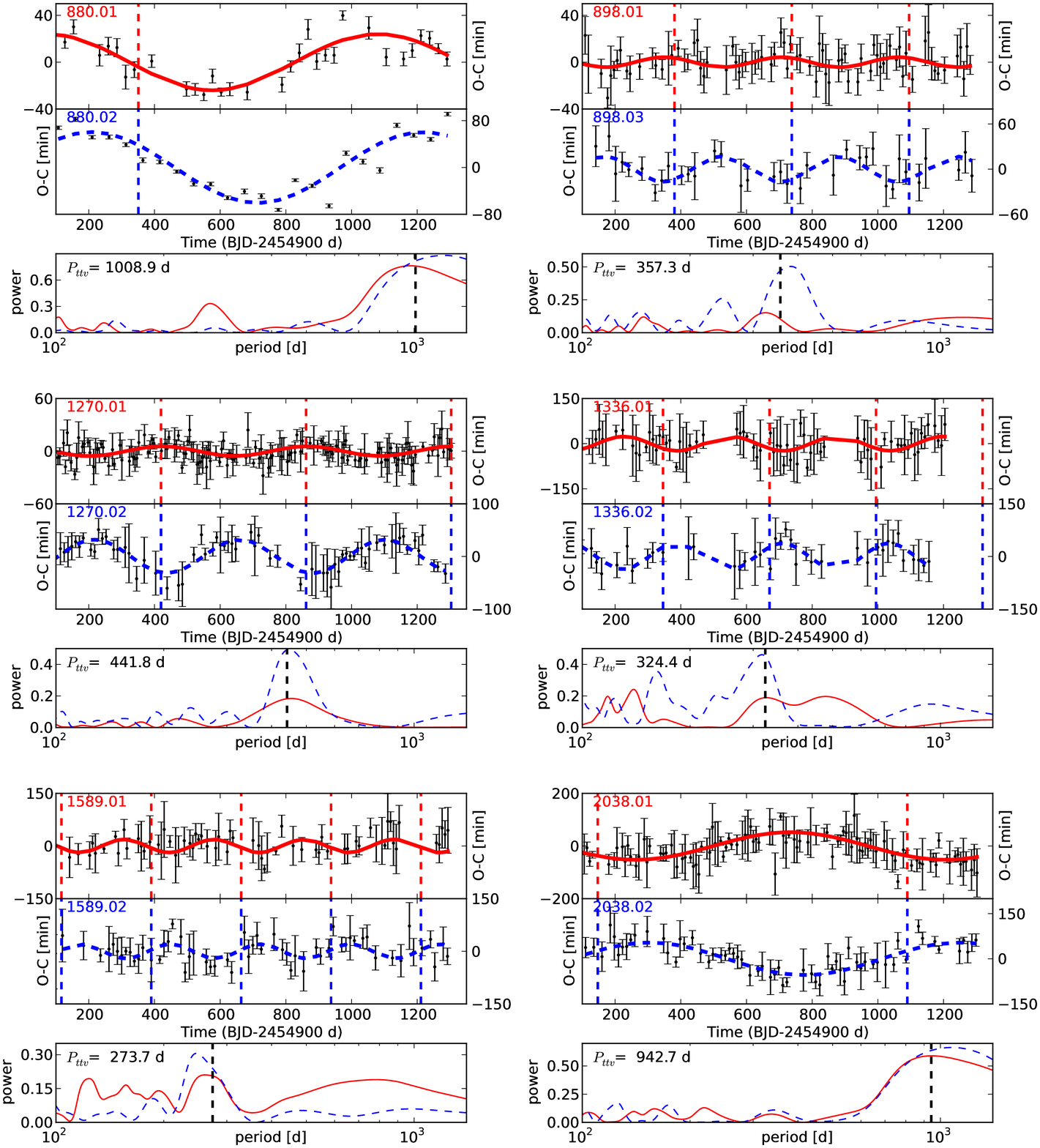}
\caption{\rev{Similar to Fig.\ref{fig_ttv1}, but for another six KOI pairs. }}
\label{fig_ttv2}
   \end{center}
\end{figure}

\begin{figure}
\begin{center}
\includegraphics[width=0.45\textwidth]{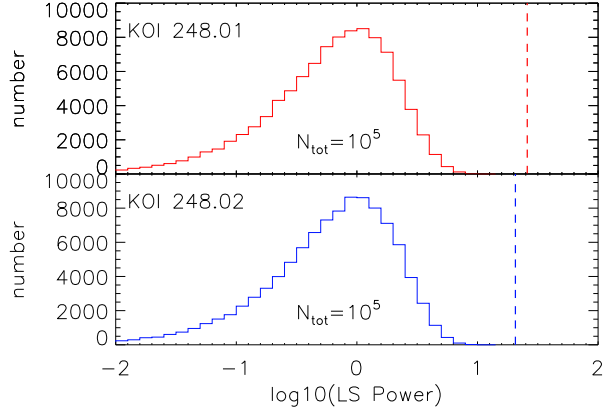}
\caption{\rev{Study of the false alarm probability (FAP) of the observed TTVs for KOI pair 248.01 and 248.02. The histograms show the LS power (at the super-period, $P^{j}$) distribution of $10^{5}$ random realizations of the originally observed TTV data, and the vertical dashed line marks the power of the originally observed TTV. None of the random realization could produce a pair of TTV with signal power larger than the originally observed one, indicating a very low FAP ($<10^{-5}$) for the originally observed TTV pair. }  }
\label{fig_fap}
   \end{center}
\end{figure}

\begin{figure}
\begin{center}
\includegraphics[width=\textwidth]{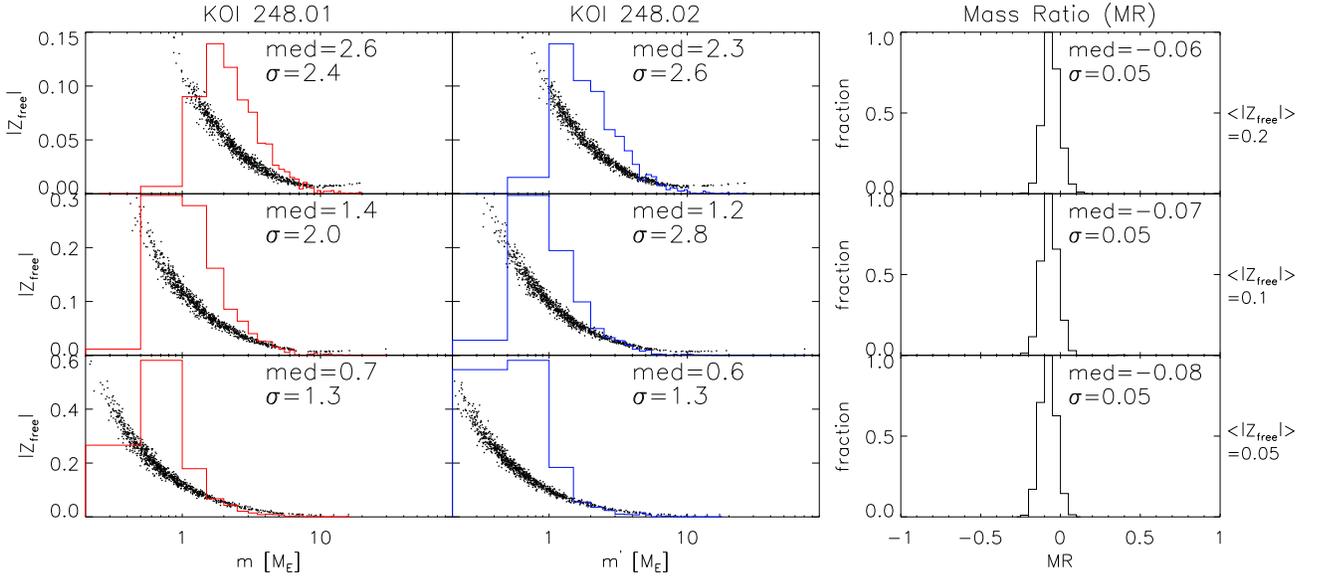}
\caption{\rev{Constraints on masses (1st and 2nd columns) and mass ratio ($\rm MR={\rm log}_{10}(m^{'}/m)$, 3rd column) of KOI pair 248.01 and 248.02 from observed TTV phases and amplitudes by performing Monte-Carlo simulations if assuming a Rayleigh distribution to the prior of free eccentricity modulus, $|Z_{\rm free}|$, with mean equal to 0.05 (top), 0.1(middle) and 0.2(bottom).  The histogram in each panel shows the distribution on the horizontal axis. The median (med) and standard deviation ($\sigma$) are printed at the top-right corner. We see the well known degeneracy between eccentricity and mass, and thus individual masses cannot be well constrained. However, the two planets' masses are certainly confined in the planetary range ($<25M_{J}\sim7945\, M_{\rm E}$, Schneider et al. 2011), and their ratio is relatively well constrained.} }
\label{fig_mass1}
   \end{center}
\end{figure}

\begin{figure}
\begin{center}
\includegraphics[width=\textwidth]{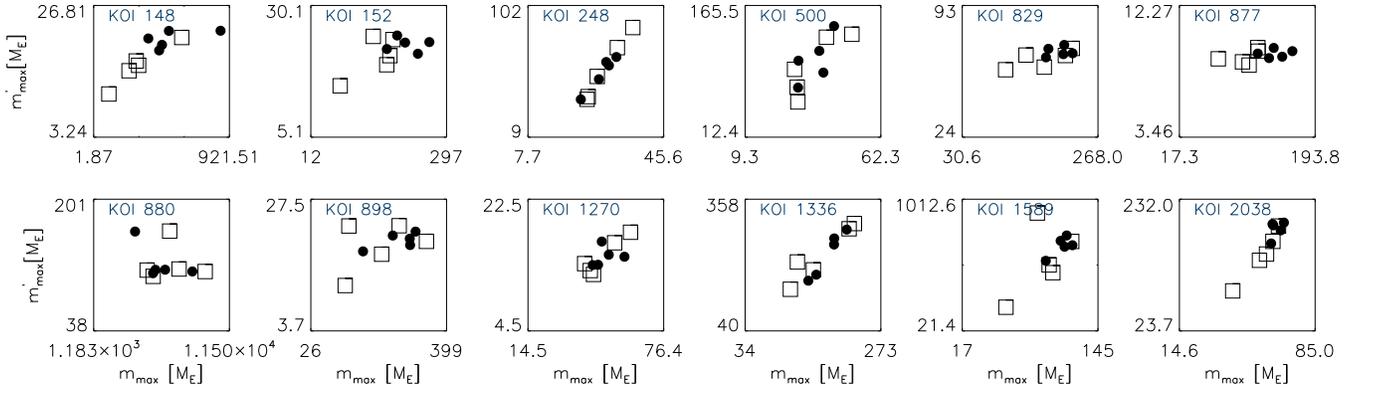}
\caption{\x{Maximum masses ($m_{\rm max}$ vs. $m^{'}_{\rm max}$)  obtained from 10 Monte Carlo simulations by setting the prior of the free eccentricity modulus, $|Z_{\rm free}|$, as a Rayleigh distribution  with a mean equal to 0.0125, 0.025, 0.05, 0.1, 0.2 (5 squares) or a uniform distribution (5 filled circles) with an upper cut at 0.025, 0.05, 0.1, 0.2, and 0.4, respectively. We see all maximum masses are within the planetary range ($<25M_{J}\sim7945\, M_{\rm E}$, Schneider et al. 2011).}}
\label{fig_mass2}
   \end{center}
\end{figure}

\begin{figure}
\begin{center}
\includegraphics[width=\textwidth]{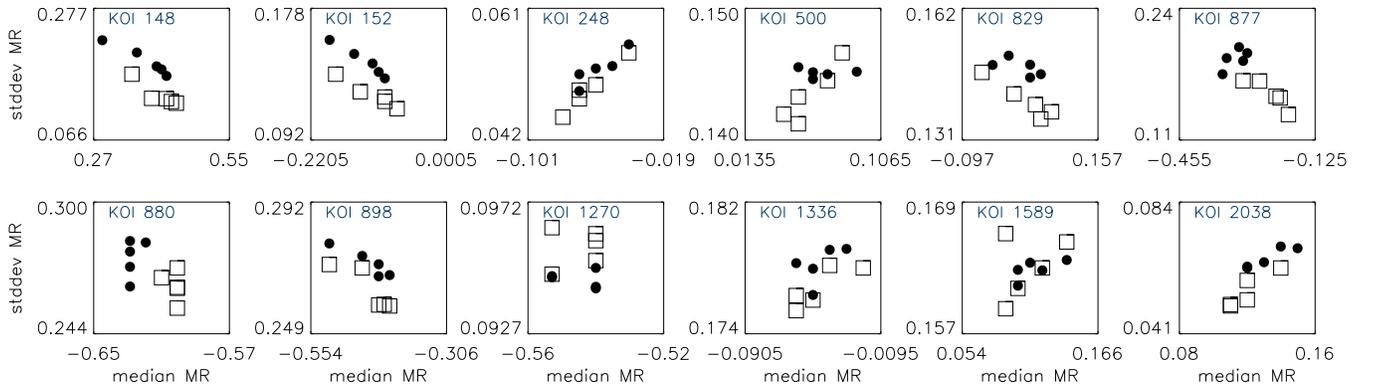}
\caption{\x{Similar to figure \ref{fig_mass2} but the horizontal and vertical axes are the median and standard deviation of mass ratio index $\rm MR={\rm log}_{10} (m^{'}/m)$.} }
\label{fig_mr}
   \end{center}
\end{figure}

\begin{figure}
\begin{center}
\includegraphics[width=0.9\textwidth]{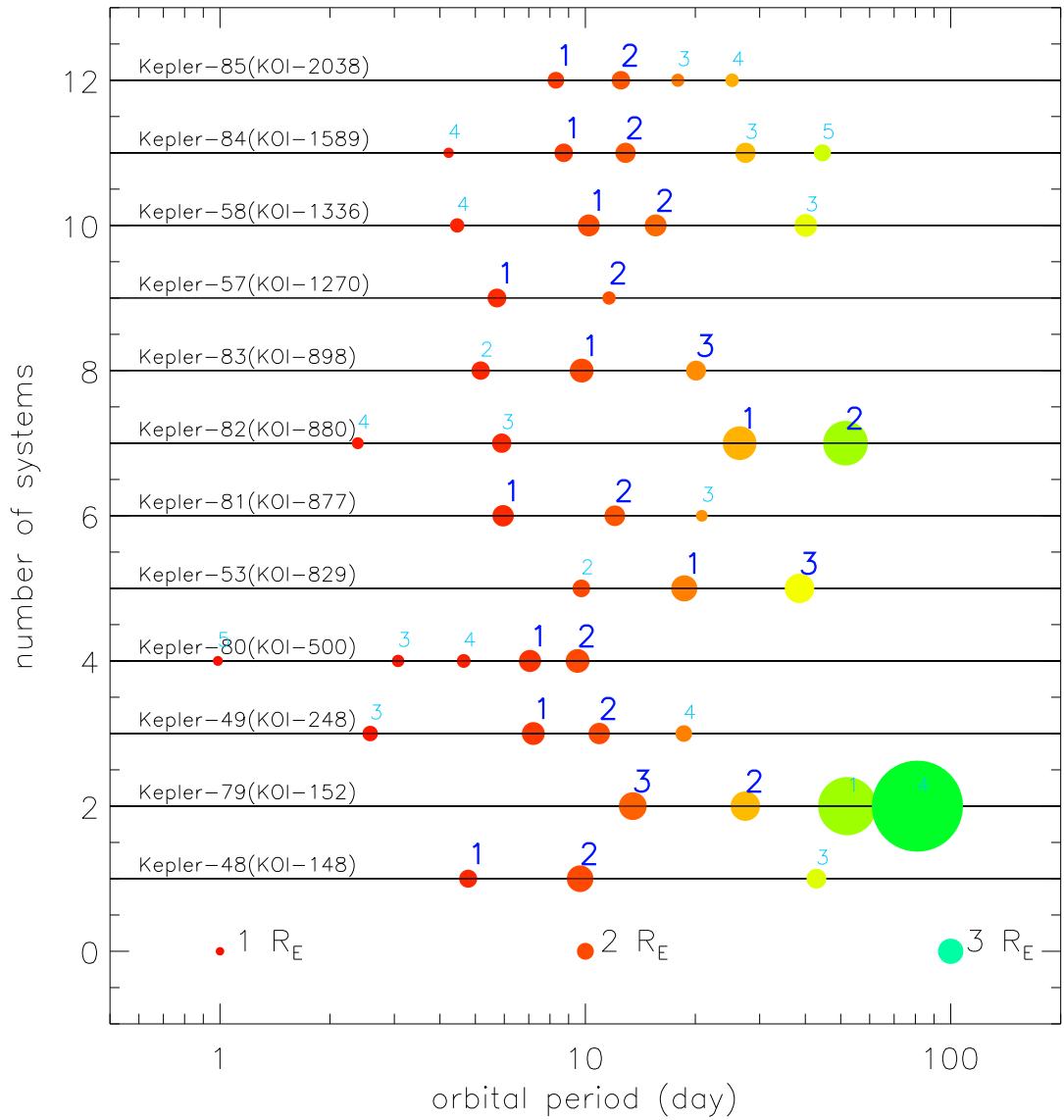}
\caption{Twelve multiple transiting systems studied in this paper. Planets and candidates are plotted and coloured in order of orbital period. The number beside each planet (in larger blue font, confirmed in this paper) and candidate (in smaller cyan font) is the KOI sequence id. }
\label{fig_sys}
   \end{center}
\end{figure}

\begin{figure}
\begin{center}
\includegraphics[width=0.45\textwidth]{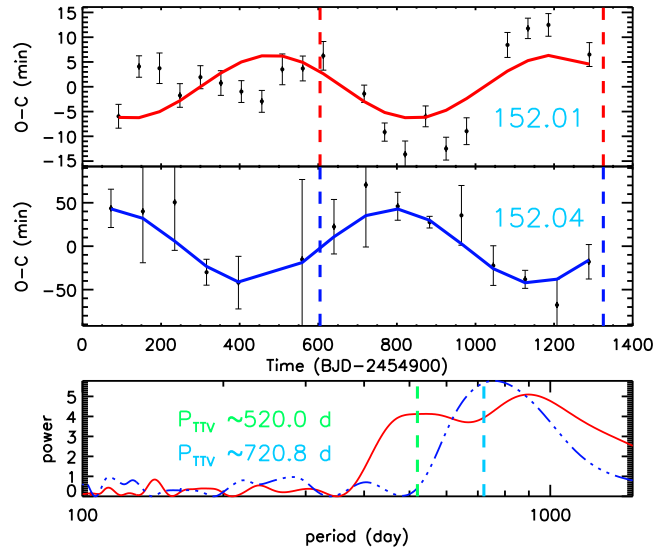}
\caption{\rev{Similar to Fig.\ref{fig_ttv1}, but for KOI pair 152.01 and 152.04. This pair is close to 3:2 MMR with $\Delta =0.038$ and theoretical TTV period of $\sim720$ d. Such a TTV mode is shown clearly on KOI-152.04 but a little bit weak on KOI-152.01. The latter also show a TTV mode on a period of $\sim 520$ d, which is due to its interaction with KOI-152.02.  Note,  due to very few TTV data points for the KOI-152.04 (because of its long orbital period), we don't remove the TTV outliers with very large error bars in this plot.}}
\label{fig_152}
   \end{center}
\end{figure}

\begin{itemize}
\item \emph{Kepler-79=KOI-152}\\
This system has four transiting objects (in period order) KOI 152.03, 152.02, 152.01 and 152.04, with orbital periods about 13.5, 27.4, 52.1, and 81.07 d and radii about 3.3, 3.5, 6.9 and 10.8 Earth radius ($R_{E}$), respectively.  The four planets form three adjacent pairs which are all near MMR. The inner pair has an orbital ratio of 2.037, thus near 2:1 MMR with $\Delta = 0.018$ and theoretical TTV period of $\sim851$ d.  An expected sinusoidal TTV pair which leads to confirmation of these two planets is shown in figure \ref{fig_ttv1}.  The middle pair has a period ratio about 1.9, thus near 2:1 MMR with $\Delta\sim0.05$ and theoretical TTV period of $\sim526$ d. It is not unexpected that such a TTV mode is not significant on KOI-152.02, as the planet is dominated by the interaction of the inner pair, which is much closer to MMR (smaller $\Delta$) and thus with stronger interactions. Nevertheless, the TTV mode is indeed seen in KOI-152.01 (Fig.\ref{fig_152}), and moreover, there is another TTV mode with longer period showing in its TTV. This longer mode is expected and it is caused as it's involved in the interaction of the outer pair, which has a period ratio of 1.556, thus near 3:2 MMR with $\Delta =0.038$ and theoretical TTV period of $\sim720$ d. As also expected, KOI-152.04 also show such a mode in its TTV (Fig.\ref{fig_152}). All these expected TTV behaviours strongly suggest that KOI-152 is a particular interesting system with 4 planets interacting in a MMR chain (1:2:4:6). However, due to the relatively few TTV measurements (especially for KOI-152.01), we conservatively prefer not to claim the confirmation of KOI 152.01 and 152.04 in this paper until more data is available in the future.

\item \emph{Kepler-80=KOI-500}\\
This system has five transiting objects (in period order) KOI 500.05, 500.03. 500.04, 500.01 and 500.02 with orbital periods about 0.99, 3.07, 4.65, 7.05 and 9.52 d and radii about 1.2, 1.5, 1.6, 2.6 and 2.8 $R_{E}$, respectively. The outer two, KOI 500.01 and 500.02 are near a 4:3 MMR with $\Delta=0.012$ and show the expected TTV profile (Fig.\ref{fig_ttv1}). 
For the inner three smaller objects,  KOI 500.05 is not involved in any MMR, and KOI 500.03 and 500.04 are close to 3:2 MMR. In addition, KOI 500.04 and 500.01 are close to 3:2 MMR, and KOI 500.04 and 500.02 are close to 2:1 MMR.  In total, there are 4 pairs near MMR in the system of KOI-500, and interestingly, all of them have nearly identical super-period, $P^{j}\sim$191-192 d. However, due to the relatively small size and thus low signal noise ratio on the TTV, we are not able to confirm the inner three ones using the method of this paper.  Future studies are needed to further unveil such a peculiar system \footnote{During the revision process of this paper, another publication \citep{Rag12} investigated in detail the properties of KOI-500 system.}.

\item \emph{Kepler-82 = KOI-880}\\
This system has four transiting objects (in period order) KOI 880.04, 880.03, 880.01 and 880.02 with orbital periods about 2.38, 5.90, 26.44 and 51.53 d and radii about 1.4, 2.3, 4.0 and 5.3 $R_{E}$, respectively. The inner two smaller objects are not involved in any first or second order of MMR and thus do not show any interesting TTV signal. The outer two, KOI 880.01 and 880.02, are much larger and near 2:1 MMR with $\Delta=-0.026$, which show expected sinusoidal TTV (Fig.\ref{fig_ttv2}). Unlike most other KOI pairs which show anti-correlated TTV with phases difference close to $\pi$, their TTV are more or less correlated with phase difference close to 0. Such TTV phases, though unusual,  are indeed allowed and predicted from theory (Eqn.\ref{vv}), which requires the system to have a free eccentricity $E_{\rm free}\sim \Delta$.   
Furthermore, the outer pair seems to have a mass ratio of $m/m^{'}\sim10^{0.6}\sim4$ from the TTV constraint (Fig.\ref{fig_mr}), which would lead to a very large density ratio of $4\times(5.35/4)^{3}\sim10$ (the current  record holder is Kepler 36 b/c with a density ratio about 8). Nevertheless, one should note that the above estimate is based on a small number of TTV measurements and it is still subject to large uncertainty in the mass ratio (a fact of $\sim2$ according to Fig.\ref{fig_mr}). Future studies with more observations (e.g., longer TTV data) are needed to further accurately characterize this interesting system.
\end{itemize}

\subsection{Future prospects} 
With more and longer data, TTV can place tighter constraints on the mass and dynamical evolution of these already confirmed planets, and will continue to confirm more, pushing the limit of confirmation to planets with smaller sizes and longer periods (e.g., KOI 152.01 and 152.04). As the confirmed sample grows and more interesting systems (e.g., KOI-500) are discovered, it would be interesting to perform some studies either on all the systems to obtain some statistical properties of different populations (e.g., \citet{WL12}) or  on some individual peculiar systems to explore their properties and formation history (e.g., \citet{Rag12}). All of these will improve our knowledge of exoplanets and deepen our understanding of planet formation and evolution.    

\acknowledgments 
JWX thanks the referee for a constructive review report, Yanqin Wu and Yoram Lithwick for helpful conversations on TTV data analysis,  Ji Wang for assistance in checking the centroid offset of the transit targets, Jason Steffen and Tomer Holczer for TTV comparison, Ernst De Mooij for discussion of red noise, Wolfgang Kerzendorf and Amir Hajian for installation and discussion on MCMC fitting with Python, John Dubinski for providing access to the Sunnyvale cluster at CITA on which most calculation of this work was performed and Matthew Payne for reading and revising the manuscript. JWX acknowledges support from the Ontario government, University of Toronto, the Key Development Program of Basic Research of China (973 program,  No. 2013CB834900),   National Natural Science Foundation of China (No. 10925313), 985 Project of Ministration of Education, Superiority Discipline Construction Project of Jiangsu Province. This work would not be done without the beautiful light curves produced by the Kepler team.



\clearpage
\begin{table*}[]
 \begin{center}
  \caption{Transit time measurements of 12 pair of KOIs. }
  \label{tb_ttv}
\begin{tabular}{cccc}
\hline
\hline
$^{a}$ KOI 148.01  &      Ntr=      &   208 \\ 
           n          &       t (d)                 &   et (d) \\
           0          &     57.0449257      &    0.0046222\\
           1          &     61.8420258      &   0.0097333\\         
           2          &     66.6172867      &   0.0063346\\
          . . .        &     . . .                    &    . . .   \\
         260        &   1299.3457031    &   0.0094695\\
 KOI 148.02  &      Ntr=      &   91 \\ 
           n          &       t (d)                 &   et (d) \\
           0          &       58.3351860     &  0.0042707\\
           1          &       68.0150299     &  0.0028628\\ 
           2          &       77.6862793     &  0.0024689\\
          . . .        &     . . .                     &    . . .   \\
         128        &   1296.6087646     &  0.0022664\\
 KOI 152.03  &      Ntr=      &   69 \\      
           n          &       t (d)                 &   et (d) \\      
          . . .       &     . . .                  &    . . .   \\     
\hline
\hline\\
\end{tabular}\\
$^{a}$ Here $n$ is the transit sequence id, $\rm t= BJD - 2454900$ d is the transit time with its uncertainty, ${\rm et}$, and Ntr is the number of transits for each candidate. This table will be available in its entirety in a machine-readable form (also available at \url{http://www.astro.utoronto.ca/$\sim$jwxie/TTV}). A portion is shown here for guidance regarding its form and content.

\end{center}
\end{table*}

\begin{table*}
  \caption{Results of TTV Analyses for 12 Pairs of Planets}
  \label{tb_mr}
\begin{center}
\resizebox{\textwidth}{!}{%
\renewcommand{\arraystretch}{1.8}
\begin{tabular}{|rr|rrr|rrrr|rr|}
\hline
Kepler & KOI & $j$ & $\Delta$ & $P^{j}$ & $A_{ttv}$ & $A^{'}_{ttv}$ & $\phi_{ttv}$ & $\phi^{'}_{ttv}$ & $\rm FAP_1$ & $\rm FAP_2$ \\
-   &-   &   - & -      & d    & d       & d             & deg          & deg            & -           & -  \\
\hline
48 b-c&148{.01}-{02}&       2&  0.012&   391.9&$    0.0030_{-    0.0005}^{+    0.0005}$&$    0.0016_{-    0.0003}^{+    0.0003}$&$   -14.5_{-    11.5}^{+    11.5}$&$   190.2_{-    15.0}^{+    15.0}$&$< 10^{-5}$&$< 10^{-4}$\\
79 b-c&152{.03}-{02}&       2&  0.016&   853.4&$    0.0060_{-    0.0012}^{+    0.0012}$&$    0.0112_{-    0.0016}^{+    0.0016}$&$   -26.6_{-    12.2}^{+    12.2}$&$   140.2_{-    10.2}^{+    10.2}$&$< 10^{-5}$&$< 10^{-4}$\\
49 b-c&248{.01}-{02}&       3&  0.010&   367.6&$    0.0078_{-    0.0006}^{+    0.0006}$&$    0.0127_{-    0.0009}^{+    0.0009}$&$    31.9_{-     3.8}^{+     3.8}$&$   205.6_{-     4.4}^{+     4.4}$&$< 10^{-5}$&$< 10^{-4}$\\
80 b-c&500{.01}-{02}&       4&  0.012&   191.3&$    0.0042_{-    0.0008}^{+    0.0008}$&$    0.0048_{-    0.0012}^{+    0.0012}$&$    -6.3_{-    11.1}^{+    11.1}$&$   168.1_{-    11.3}^{+    11.3}$&$< 10^{-5}$&$< 10^{-4}$\\
53 b-c&829{.01}-{03}&       2&  0.034&   571.1&$    0.0125_{-    0.0027}^{+    0.0027}$&$    0.0158_{-    0.0029}^{+    0.0029}$&$   -39.3_{-    10.7}^{+    10.7}$&$   112.8_{-    11.0}^{+    11.0}$&$< 10^{-5}$&$< 10^{-4}$\\
81 b-c&877{.01}-{02}&       2&  0.011&   551.1&$    0.0025_{-    0.0005}^{+    0.0005}$&$    0.0059_{-    0.0010}^{+    0.0010}$&$    26.8_{-    11.9}^{+    11.9}$&$   250.5_{-     8.0}^{+     8.0}$&$< 10^{-5}$&$< 10^{-4}$\\
82 b-c&880{.01}-{02}&       2& -0.026&  1008.9&$    0.0167_{-    0.0010}^{+    0.0010}$&$    0.0421_{-    0.0006}^{+    0.0006}$&$     9.9_{-     9.6}^{+     9.6}$&$   320.5_{-     8.3}^{+     8.3}$&$< 10^{-5}$&$< 10^{-4}$\\
83 b-c&898{.01}-{03}&       2&  0.028&   357.3&$    0.0030_{-    0.0013}^{+    0.0013}$&$    0.0117_{-    0.0028}^{+    0.0028}$&$    58.1_{-    24.2}^{+    24.2}$&$   234.5_{-    14.3}^{+    14.3}$&   0.00009&    0.0005\\
57 b-c&1270{.01}-{02}&       2&  0.013&   441.8&$    0.0038_{-    0.0007}^{+    0.0007}$&$    0.0221_{-    0.0027}^{+    0.0027}$&$    92.9_{-    10.6}^{+    10.6}$&$   284.8_{-     4.8}^{+     4.8}$&$< 10^{-5}$&$< 10^{-4}$\\
58 b-c&1336{.01}-{02}&       3&  0.016&   324.4&$    0.0165_{-    0.0045}^{+    0.0045}$&$    0.0259_{-    0.0063}^{+    0.0063}$&$   -42.4_{-    15.0}^{+    15.0}$&$   133.5_{-    11.5}^{+    11.5}$&$< 10^{-5}$&$< 10^{-4}$\\
84 b-c&1589{.01}-{02}&       3& -0.016&   273.7&$    0.0129_{-    0.0034}^{+    0.0034}$&$    0.0140_{-    0.0031}^{+    0.0031}$&$    18.7_{-    15.6}^{+    15.6}$&$   192.4_{-    15.9}^{+    15.9}$&   0.00020&    0.0088\\
85 b-c&2038{.01}-{02}&       3&  0.004&   942.7&$    0.0362_{-    0.0029}^{+    0.0029}$&$    0.0374_{-    0.0030}^{+    0.0030}$&$   -45.9_{-     4.7}^{+     4.7}$&$   150.2_{-     5.1}^{+     5.1}$&$< 10^{-5}$&$< 10^{-4}$\\
\hline
\end{tabular}} \\
\end{center}

\end{table*}

\begin{table*}
  \caption{Key Properties of Planets and Stars of the 12 Systems.}
  \label{tb_ttv}
\begin{center}
\renewcommand{\arraystretch}{1.8}
\resizebox{\textwidth}{!}{%
\begin{tabular}{|rr|rrrrrrrr|rrrr|}
\hline
Kepler &KOI & $P$$^{a}$ & $P^{'}$$^{a}$ & $R$$^{b}$  & $R^{'}$$^{b}$ & $m_{\rm nom}$$^{c}$ & $m^{'}_{\rm nom}$$^{c}$ & $m_{\rm max}$$^{d}$ & $m^{'}_{\rm max}$$^{d}$ & $M_{\star}$$^{e}$ & log(g)$^{f}$ & $R_{\star}$$^{f}$ &  $\xi$$^{g}$ \\
- & -  & d   &  d      &$R_{\rm E}$ & $R_{\rm E}$   & $M_{\rm E}$         & $M_{\rm E}$             & $M_{\rm E}$         & $M_{\rm E}$             & $M_{\odot}$    &-  & $R_{\odot}$       &  -           \\
\hline
48 b-c&148{.01}-{02}&   4.778&   9.674&$2.14_{-0.12}^{+0.12}$&$3.14_{-0.18}^{+0.18}$&$     9.0_{-     6.4}^{+    12.5}$&$     9.4_{-     6.8}^{+    12.9}$&   614.3&    17.9&$  0.88_{ -0.20}^{+  0.26}$&$4.49_{-0.10}^{+0.10}$&$0.89_{-0.05}^{+0.05}$&1.03\\
79 b-c&152{.03}-{02}&  13.485&  27.402&$2.59_{-0.82}^{+0.82}$&$2.77_{-0.88}^{+0.88}$&$    34.7_{-    12.4}^{+    85.2}$&$    10.7_{-     3.8}^{+    26.9}$&   197.9&    20.0&$  1.10_{ -0.70}^{+  1.63}$&$4.42_{-0.30}^{+0.30}$&$1.10_{-0.35}^{+0.35}$&0.94\\
49 b-c&248{.01}-{02}&   7.204&  10.913&$2.72_{-0.12}^{+0.12}$&$2.55_{-0.13}^{+0.13}$&$     7.7_{-     3.8}^{+    15.6}$&$     7.8_{-     3.8}^{+    15.7}$&    30.4&    67.9&$  0.55_{ -0.27}^{+  0.56}$&$4.74_{-0.30}^{+0.30}$&$0.52_{-0.01}^{+0.01}$&1.20\\
80 b-c&500{.01}-{02}&   7.053&   9.522&$2.64_{-0.11}^{+0.11}$&$2.79_{-0.13}^{+0.13}$&$     5.7_{-     4.2}^{+     7.5}$&$     7.1_{-     5.5}^{+     8.9}$&    41.5&   110.3&$  0.72_{ -0.10}^{+  0.11}$&$4.67_{-0.06}^{+0.06}$&$0.65_{-0.02}^{+0.02}$&1.12\\
53 b-c&829{.01}-{03}&  18.649&  38.558&$2.89_{-0.17}^{+0.17}$&$3.17_{-0.19}^{+0.19}$&$    72.5_{-    48.0}^{+   107.3}$&$    33.0_{-    21.7}^{+    50.2}$&   178.6&    61.9&$  1.07_{ -0.32}^{+  0.46}$&$4.40_{-0.15}^{+0.15}$&$1.09_{-0.05}^{+0.05}$&1.14\\
81 b-c&877{.01}-{02}&   5.955&  12.040&$2.42_{-0.38}^{+0.38}$&$2.37_{-0.37}^{+0.37}$&$    16.5_{-     9.9}^{+    27.3}$&$     4.0_{-     2.4}^{+     6.7}$&   129.2&     8.2&$  0.64_{ -0.23}^{+  0.38}$&$4.70_{-0.20}^{+0.20}$&$0.59_{-0.03}^{+0.03}$&1.09\\
82 b-c&880{.01}-{02}&  26.444&  51.538&$4.00_{-1.82}^{+1.82}$&$5.35_{-2.44}^{+2.44}$&$    86.1_{-    22.1}^{+   248.0}$&$    19.0_{-     4.8}^{+    55.0}$&  7663.6&   133.8&$  0.85_{ -0.63}^{+  1.66}$&$4.49_{-0.30}^{+0.30}$&$0.90_{-0.41}^{+0.41}$&0.78\\
83 b-c&898{.01}-{03}&   9.770&  20.090&$2.83_{-0.41}^{+0.41}$&$2.36_{-0.35}^{+0.35}$&$    51.2_{-    29.7}^{+    86.8}$&$     7.5_{-     3.6}^{+    13.8}$&   265.8&    18.3&$  0.66_{ -0.25}^{+  0.41}$&$4.69_{-0.20}^{+0.20}$&$0.61_{-0.03}^{+0.03}$&0.93\\
57 b-c&1270{.01}-{02}&   5.729&  11.609&$2.19_{-0.95}^{+0.95}$&$1.55_{-0.67}^{+0.67}$&$    95.0_{-    30.3}^{+   257.8}$&$     9.5_{-     3.0}^{+    25.8}$&    50.9&    15.0&$  0.76_{ -0.51}^{+  1.25}$&$4.62_{-0.30}^{+0.30}$&$0.74_{-0.27}^{+0.27}$&0.83\\
58 b-c&1336{.01}-{02}&  10.219&  15.573&$2.78_{-1.18}^{+1.18}$&$2.86_{-1.21}^{+1.21}$&$    31.9_{-     8.6}^{+    91.5}$&$    32.3_{-     8.7}^{+    93.9}$&   182.1&   238.8&$  0.97_{ -0.70}^{+  1.65}$&$4.38_{-0.30}^{+0.30}$&$1.09_{-0.46}^{+0.46}$&0.99\\
84 b-c&1589{.01}-{02}&   8.726&  12.883&$2.23_{-0.10}^{+0.10}$&$2.36_{-0.11}^{+0.11}$&$    21.3_{-    13.8}^{+    32.1}$&$    30.5_{-    19.1}^{+    47.1}$&    96.7&   675.1&$  1.00_{ -0.29}^{+  0.42}$&$4.13_{-0.15}^{+0.15}$&$1.43_{-0.05}^{+0.05}$&1.13\\
85 b-c&2038{.01}-{02}&   8.306&  12.513&$1.97_{-0.10}^{+0.10}$&$2.18_{-0.10}^{+0.10}$&$    15.5_{-    10.7}^{+    22.4}$&$    24.1_{-    16.6}^{+    34.8}$&    56.7&   154.7&$  0.92_{ -0.27}^{+  0.40}$&$4.22_{-0.15}^{+0.15}$&$1.24_{-0.05}^{+0.05}$&1.01\\
\hline
\end{tabular}} \\
\end{center}
$^{a}$ Their uncertainties are all less than $10^{-3}$ d\\
$^{b}$ Their errorbars reflect the uncertainties of their transit lightcurve fittings (i.e., $R_{\rm p}/R_{\star}$) and stellar radii.\\
$^{c}$ Derived from Eqn.\ref{mn}. Their errorbars reflect the uncertainties of their TTV emplitudes and stellar masses.\\
$^{d}$ Obtained from Monte Carlo simulations as described in section 3.3. \\
$^{e}$ Derived from $log(g)$ and $R_{\star}$.\\
$^{f}$ For KOI = 829, 1589 and 2038, their stellar parameters are adopted from \citet{Eve13}, otherwise, they are adopted from the NASA exoplanet archive (http://exoplanetarchive.ipac.caltech.edu). \\
$^{g}$ The normalized transits duration ratio, $\xi=(T_{\rm dur}/T_{\rm dur}^{'})(P^{'}/P)^{1/3}$ \citep{Fab12b}. \\

\end{table*}

\clearpage

\clearpage

\clearpage

\clearpage
\clearpage


\clearpage


\newpage
\begin{thebibliography}{99}

\bibitem[Agol et al.(2005)]{Ago05} Agol, E., Steffen, J., 
Sari, R., \& Clarkson, W.\ 2005, \mnras, 359, 567 

\bibitem[Baraffe et 
al.(2004)]{Bar04} Baraffe, I., Selsis, F., Chabrier, G., et al.\ 2004, \aap, 419, L13

\bibitem[Batalha et al.(2011)]{Bat11} Batalha, N.~M., 
Borucki, W.~J., Bryson, S.~T., et al.\ 2011, \apj, 729, 27 

\bibitem[Batalha et al.(2012)]{Bat12} Batalha, N.~M., Rowe, 
J.~F., Bryson, S.~T., et al.\ 2012, arXiv:1202.5852

\bibitem[Batygin 
\& Morbidelli(2012)]{BM12} Batygin, K., \& Morbidelli, A.\ 2012, arXiv:1204.2791

\bibitem[Burke et al.(2013)]{Bur13} Burke, C.~J., Bryson, S., 
Christiansen, J., et al.\ 2013, American Astronomical Society Meeting 
Abstracts, 221, \#216.02

\bibitem[Burnham \& Anderson (2002)]{BA02}Burnham, K. P.; Anderson, D. R. (2002), Model Selection and Multimodel Inference: A Practical Information-Theoretic Approach (2nd ed.), Springer-Verlag, ISBN 0-387-95364-7

\bibitem[Carter et al.(2008)]{Car08} Carter, J.~A., Yee, 
J.~C., Eastman, J., Gaudi, B.~S., \& Winn, J.~N.\ 2008, \apj, 689, 499

\bibitem[Cochran et al.(2011)]{Coc11} Cochran, W.~D., 
Fabrycky, D.~C., Torres, G., et al.\ 2011, \apjs, 197, 7 

\bibitem[de Mooij et 
al.(2011)]{Dem11} de Mooij, E.~J.~W., de Kok, R.~J., Nefs, S.~V., \& Snellen, I.~A.~G.\ 2011, \aap, 528, A49

\bibitem[Everett et al.(2013)]{Eve13} Everett, M.~E., Howell, 
S.~B., Silva, D.~R., \& Szkody, P.\ 2013, \apj, 771, 107 

\bibitem[Fabrycky et al.(2012a)]{Fab12a} Fabrycky, D.~C., Ford, 
E.~B., Steffen, J.~H., et al.\ 2012a, \apj, 750, 114 

\bibitem[Fabrycky et al.(2012b)]{Fab12b} Fabrycky, D.~C., 
Lissauer, J.~J., Ragozzine, D., et al.\ 2012b, arXiv:1202.6328

\bibitem[Ford(2005)]{For05} Ford, E.~B.\ 2005, \aj, 129, 1706

\bibitem[Ford et al.(2012a)]{For12a} Ford, E.~B., Fabrycky, 
D.~C., Steffen, J.~H., et al.\ 2012a, \apj, 750, 113

\bibitem[Ford et al.(2012b)]{For12b} Ford, E.~B., Ragozzine, 
D., Rowe, J.~F., et al.\ 2012b, \apj, 756, 185

\bibitem[Fressin et al.(2011)]{Fre11} Fressin, F., Torres, 
G., D{\'e}sert, J.-M., et al.\ 2011, \apjs, 197, 5 

\bibitem[Gelman \& Rubin (1992)]{GR92}Gelman, A., \& Rubin, D. B. 1992, Stat. Sci., 7, 457

\bibitem[Gillon et 
al.(2007)]{Gil07} Gillon, M., Demory, B.-O., Barman, T., et al.\ 2007, \aap, 471, L51

\bibitem[Hansen 
\& Murray(2012)]{HM12} Hansen, B.~M.~S., \& Murray, N.\ 2012, \apj, 751, 158

\bibitem[Holman 
\& Murray(2005)]{HM05} Holman, M.~J., \& Murray, N.~W.\ 2005, Science, 307, 1288

\bibitem[Holman et al.(2010)]{Hol10} Holman, M.~J., Fabrycky, 
D.~C., Ragozzine, D., et al.\ 2010, Science, 330, 51

\bibitem[Huang et al.(2012)]{Hua12} Huang, X., Bakos, 
G.~{\'A}., \& Hartman, J.~D.\ 2012, arXiv:1205.6492 

\bibitem[Kipping(2010)]{Kip10} Kipping, D.~M.\ 2010, \mnras, 
408, 1758

\bibitem[Lee 
\& Peale(2002)]{LP02} Lee, M.~H., \& Peale, S.~J.\ 2002, \apj, 567, 596

\bibitem[Lissauer et al.(2011b)]{Lis11} Lissauer, J.~J., 
Fabrycky, D.~C., Ford, E.~B., et al.\ 2011, \nat, 470, 53

\bibitem[Lissauer et al.(2011a)]{Lis11a} Lissauer, J.~J., 
Ragozzine, D., Fabrycky, D.~C., et al.\ 2011, \apjs, 197, 8 

\bibitem[Lissauer et al.(2013)]{Lis13} Lissauer, J.~J., 
Jontof-Hutter, D., Rowe, J.~F., et al.\ 2013, arXiv:1303.0227

\bibitem[Lithwick 
\& Wu(2012)]{LW12} Lithwick, Y., \& Wu, Y.\ 2012, \apjl, 756, L11

\bibitem[Lithwick et al.(2012)]{Lit12} Lithwick, Y., Xie, J., 
\& Wu, Y.\ 2012, arXiv:1207.4192

\bibitem[Mandel 
\& Agol(2002)]{MA02} Mandel, K., \& Agol, E.\ 2002, \apjl, 580, L171

\bibitem[Markwardt(2009)]{Mar09} Markwardt, C.~B.\ 2009, 
Astronomical Data Analysis Software and Systems XVIII, 411, 251 

\bibitem[Mazeh et al.(2013)]{Maz13} Mazeh, T., Nachmani, G., 
Holczer, T., et al.\ 2013, arXiv:1301.5499

\bibitem[Nesvorn{\'y} et al.(2012)]{Nes12} Nesvorn{\'y}, D., 
Kipping, D.~M., Buchhave, L.~A., et al.\ 2012, Science, 336, 1133

\bibitem[Nesvorny et al.(2013)]{Nes13} Nesvorny, D., Kipping, 
D., Terrell, D., et al.\ 2013, arXiv:1304.4283

\bibitem[Ofir 
\& Dreizler(2012)]{OD12} Ofir, A., \& Dreizler, S.\ 2012, arXiv:1206.5347

\bibitem[Patil et al.(2010)]{Pat10}Patil, A., D. Huard and C.J. Fonnesbeck. 2010. PyMC: Bayesian
Stochastic Modelling in Python. Journal of Statistical Software, 35(4), pp. 1-81

\bibitem[Pont et al.(2006)]{Pon06} Pont, F., Zucker, S., 
\& Queloz, D.\ 2006, \mnras, 373, 231

\bibitem[Press et al.(1992)]{Pre92} Press, W. H., Teukolsky, S. A., Vetterling, W. T., \& Flannery, B. P. 1992, Cambridge: University Press, Ñc1992, 2nd ed.,

\bibitem[Ragozzine 
\& Kepler Team(2012)]{Rag12} Ragozzine, D., \& Kepler Team 2012, AAS/Division for Planetary Sciences Meeting Abstracts, 44, \#200.04 

\bibitem[Scargle(1982)]{Sca82} Scargle, J.~D.\ 1982, \apj, 
263, 835 

\bibitem[Steffen et al.(2012)]{Ste12} Steffen, J.~H., 
Fabrycky, D.~C., Ford, E.~B., et al.\ 2012, \mnras, 421, 2342

\bibitem[Steffen et al.(2013)]{Ste13} Steffen, J.~H., 
Fabrycky, D.~C., Agol, E., et al.\ 2013, \mnras, 428, 1077

\bibitem[Terquem 
\& Papaloizou(2007)]{TP07} Terquem, C., \& Papaloizou, J.~C.~B.\ 2007, \apj, 654, 1110

\bibitem[Veras et al.(2011)]{Ver11} Veras, D., Ford, E.~B., 
\& Payne, M.~J.\ 2011, \apj, 727, 74

\bibitem[Wu 
\& Lithwick(2012)]{WL12} Wu, Y., \& Lithwick, Y.\ 2012, arXiv:1210.7810 

\bibitem[Zechmeister \& K{\"u}rster(2009)]{ZK09} Zechmeister, M., \& K{\"u}rster, M.\ 2009, \aap, 496, 577

\end{thebibliography}
\end{document}